# Magnetic Quantum Critical Point and Superconductivity in UPt$_3$ Doped with Pd

A. de Visser[1], M. J. Graf[2], P. Estrela[1], A. Amato[3], C. Baines[3], D. Andreica[4], F.N. Gygax[4], and A. Schenck[4]

[1]*Van der Waals-Zeeman Institute, University of Amsterdam, 1018 XE Amsterdam, The Netherlands*
[2]*Department of Physics, Boston College, Chestnut Hill, MA 02467, USA*
[3]*Paul Scherrer Institute, CH-5232 Villigen, Switzerland*
[4]*Institute for Particle Physics, ETH Zürich, PSI, CH-5232 Villigen, Switzerland*
(Received 30 March 2000)

Transverse-field muon spin relaxation measurements have been carried out on the heavy-fermion superconductor UPt$_3$ doped with small amounts of Pd. We find that the critical Pd concentration for the emergence of the large-moment antiferromagnetic phase is ~0.6 at.%Pd. At the same Pd content, superconductivity is completely suppressed. The existence of a magnetic quantum critical point in the phase diagram, which coincides with the critical point for superconductivity, provides evidence for ferromagnetic spin-fluctuation mediated odd-parity superconductivity, which competes with antiferromagnetic order.

For more than a decade now it has been recognized that superconductivity (SC) and magnetism are intimately related in strongly correlated systems, such as the high-$T_c$ cuprates, heavy-fermion materials and organic superconductors [1]. One of the key issues is to identify the nature of the attractive interaction for Cooper pairing. In conventional s-wave superconductors Cooper pairing is mediated by phonons. In strongly correlated electron systems magnetic interactions suppress s-wave SC, and therefore it has been proposed that SC is unconventional and mediated by spin fluctuations [2]. Compelling evidence for spin-fluctuation mediated SC [3] has recently been obtained for the magnetically ordered heavy-fermion materials CePd$_2$Si$_2$ and CeIn$_3$. By tuning these materials towards a magnetic quantum critical point (the Néel temperature $T_N \rightarrow 0$ K), by the application of mechanical pressure, a SC phase appeared.

The heavy-fermion superconductor UPt$_3$ ($T_c \sim 0.5$ K) has become an exemplary system to study unconventional SC [1,4]. Because of the unusual coexistence of SC and ferromagnetic (FM) spin fluctuations, the latter signaled by a pronounced $T^3 \ln T$ contribution to the low-temperature ($T$) specific heat [5,6], it has been argued that UPt$_3$ is an odd-parity spin-fluctuation mediated superconductor [6,7], in close analogy with superfluidity in $^3$He. Recent NMR measurements [8], as well as alloying experiments [9], provide solid evidence for an odd parity SC state. The thermodynamic properties and multicomponent SC phase diagram can only be explained by Ginzburg-Landau models, based on an unconventional SC order parameter [4,10]. Much attention has been devoted to models where a symmetry-breaking field (SBF) lifts the (spin) degeneracy of the 2D (or 1D) order parameter, which results in a splitting $\Delta T_c = T_c^+ - T_c^-$ of the SC phase transition [10]. Experimental evidence [11] has been put forward that the SBF is provided by *small-moment antiferromagnetism* (SMAF) which sets in at $T_{N,SMAF} \sim 6$ K [12]. This established a clear coupling between magnetism and SC in UPt$_3$.

The nature of the SMAF state itself is the subject of lively debate. It has been observed convincingly through neutron [11,12] and magnetic x-ray [13] scattering only. The ordered moment, $m = 0.02$ $\mu_B$/U-atom ($T \rightarrow 0$), is extremely small, which hampers its detection by standard bulk probes. However, NMR [8] and zero-field muon spin relaxation ($\mu$SR) experiments [14,15] also do not signal the small moment [16], which strongly suggests that the moment *fluctuates* at a rate larger than 10 MHz, but on a time scale which appears static to neutrons and x-rays. Therefore, $T_{N,SMAF}$ may be considered to represent a cross-over temperature, rather than being connected to a true phase transition. This is in line with the unusual quasi-linear increase of $m^2(T)$ below $T_{N,SMAF}$ [12].

One of the hallmarks of heavy-fermion materials is the proximity to a magnetic quantum critical point (QCP). In the case of UPt$_3$ pronounced anti-ferromagnetic (AF) phase transitions can readily be induced by chemical alloying, e.g. by substituting small amounts of Pt by Pd [17] or U by Th [18]. In the U(Pt$_{1-x}$Pd$_x$)$_3$ pseudobinaries [19], AF order of the spin-density wave type has been observed in the thermal, magnetic and transport properties in the concentration range $0.02 \leq x \leq 0.08$ (see Fig.1). Neutron-diffraction experiments [20,21] show that at optimal doping ($x = 0.05$, $T_N = 6$ K) the ordered moment of the so-termed *large-moment antiferromagnetic phase* (LMAF) is substantial, $m = 0.63 \pm 0.05$ $\mu_B$/U-atom, and that the magnetic order parameter is conventional. The magnetic structure consists of a doubling of the nuclear unit cell (space group P6$_3$/mmc) along the a*-axis, with the moments pointing along a*. LMAF is also detected by local probe techniques, such as µSR [15] and NMR [22]. $T_N(x)$ follows a rather conventional Doniach-type phase diagram [23] (see Fig.1).

The SMAF phase has clearly a different signature, although the magnetic structure is identical to the one of the LMAF phase. Neutron-diffraction experiments [21] show that SMAF is robust upon alloying with Pd and persists till at least x= 0.005. The ordered moment grows upon alloying, but $T_{N,SMAF}(x)$ remains ~6 K and does not vary at these small Pd concentrations (see Fig.1). Notice that $T_{N,SMAF}$ is also insensitive to the application of pressure [11], unlike the LMAF $T_N$ [24].

All these results strongly suggest that SMAF and LMAF are different phases with a distinctly different nature. As a consequence they might also couple differently to SC. Pressure [11] and alloying experiments [25] are consistent with SMAF acting as SBF. In order to fully understand the nature of the SC phase, it is important to examine the relation (coexistence or competition) between LMAF and SC as well. Therefore, it is crucial to determine the critical Pd concentration for the emergence of the LMAF phase. In this Letter we present µSR experiments carried out on U(Pt$_{1-x}$Pd$_x$)$_3$ samples (x= 0.007, 0.008 and 0.009), which show that the Néel temperature $T_N$ for the LMAF phase is suppressed to 0 K at a Pd concentration $x_{c,af} \approx$ 0.006. Combined with our earlier results on the suppression of SC by Pd substitution [25,26], we find that $x_{c,af} \approx x_{c,sc}$ and that unconventional SC is replaced by LMAF [19]. As we will argue below, these results provide new evidence for SC mediated by *ferromagnetic* spin fluctuations.

Polycrystalline U(Pt$_{1-x}$Pd$_x$)$_3$ samples were prepared in a two-step process. First, master alloys of UPt$_3$ and U(Pt$_{0.95}$Pd$_{0.05}$)$_3$ were prepared by arc-melting stoichiometric amounts U (purity 99.98%), Pt and Pd (both with purity 99.999%) on a water-cooled copper crucible in a high-purity argon atmosphere (0.5 bar). Next samples with x= 0.007, 0.008 and 0.009 were prepared by arc-melting together appropriate amounts of the master alloys. After an annealing procedure (see Ref.15) four thin platelets (thickness 0.8 mm, area 6x10 mm$^2$) were prepared by spark-erosion and glued with General Electric varnish on a silver support, in order to cover an area of 12x20 mm$^2$, which corresponds to the total cross-section of the muon beam. Measurements of the residual resistivity on pieces cut from the annealed buttons are consistent with previous results [26], which ensures that Pd dissolves homogeneously in the UPt$_3$ matrix.

Measurements of the positive muon (µ$^+$) precession in an applied transverse field of 100 G were conducted at the low temperature µSR facility (LTF) on the πM3 beam line at the Paul Scherrer Institute. The samples were mounted on the cold-finger of a top-loading dilution refrigerator with a base temperature of 0.025 K. As we shall see, the muon spin depolarization rates are very small. Accurate determination of the rates for the samples with x= 0.007 and 0.008 was made possible by the use of a kicker device, which ensures that only one muon at a time is present in the sample and that no other muons are present in the spectrometer [27]. This so-called MORE (Muons On REquest) mode allows to extend the µSR time window to 20 µs, with virtually no accidental background, making it possible to measure relaxation rates as small as 0.001 µs$^{-1}$.

When positive muons come to rest in the sample they start to precess around the local field, $B_{loc}$, with a precession frequency $\nu_\mu = \gamma_\mu B_{loc}$ ($\gamma_\mu/2\pi$= 135.5 MHz/T is the muon gyromagnetic ratio). The internal dipolar magnetic field distribution in general leads to de-phasing of the precession frequency and consequently the signal is damped. As a first step, we have analyzed the µSR spectra using a Gaussian-damped depolarization function $P_G(t) = A_G\cos(\omega t)\exp(-\Delta^2 t^2/2)$, where $A_G$ is the asymmetry, $\omega = 2\pi\nu_\mu$ and $\Delta$ is the Gaussian damping rate. At the highest temperatures, $\Delta$ attains a T independent value of ~0.06 µs$^{-1}$, which is consistent with depolarization due to static $^{195}$Pt nuclear moments [15]. Upon lowering T, $\Delta$ rises progressively, which points to the presence of an additional source of internal dipolar magnetic fields. Improved fit results were obtained using the damped-Gauss muon spin depolarization function

$$P_{DG}(t) = A_{DG}\cos(\omega t)\exp(-\lambda_E t - \Delta^2 t^2/2) \qquad (1)$$

with $\Delta$ fixed at the observed Pt nuclear depolarization rate ~0.06 µs$^{-1}$. In eq.(1) the factor $\exp(-\lambda_E t)$ accounts for damping due to the additional magnetic signal. Because of the low damping rates and the large sample size, the asymmetry $A_{DG}$ is close to the maximum value ~0.3. In Fig.2 the T dependence of $\lambda_E$ is shown for all three samples. At the highest temperatures $\lambda_E$ is very small ~0.003 µs$^{-1}$ and essentially T independent. Upon lowering T $\lambda_E$ increases, as the additional source of magnetism emerges. The additional source of magnetism becomes stronger when the Pd concentration increases, and we associate its onset temperature with the Néel temperature $T_N$ for LMAF.

In order to extract $T_N$, we write the observed exponential damping rate as $\lambda_E = \lambda_{BG} + \lambda_{LMAF}$, where $\lambda_{BG}$ and $\lambda_{LMAF}$ are due to the background and the LMAF phase, respectively. $\lambda_{BG}$ may account for small variations of the actual depolarization rate due to Pt nuclear moments, as in the fitting procedure we used the fixed value $\Delta$= 0.06 µs$^{-1}$. The super-linear increase of $\lambda_{LMAF}$ is unusual, and can be described phenomenologically, in this limited $T$ interval, by a quasi-logarithmic increase $\lambda_{LMAF} \sim -\ln(T/T_N)$. Making use of this functional dependence and imposing $\lambda_{LMAF}=0$ for $T>T_N$, we obtain $T_N$ values of 1.23±0.10 K, 0.78±0.10 K and 0.45±0.15 K for $x$= 0.009, 0.008 and 0.007, respectively.

The assignment of the increase of $\lambda_E$ to the onset of LMAF ordering is based on an analogy with the analysis of the zero-field μSR spectra obtained in the LMAF phase for samples with higher Pd concentrations [15]. For $x \geq 0.01$ the spectra are well described by a two-component depolarization function, consisting of the standard depolarization function of a polycrystalline antiferromagnet and a Kubo-Lorentzian (KL) term, accounting for the spectral distribution of internal fields. Within this phenomenological approach it was observed that the depolarization rate of the KL function, $\lambda_{KL}$, scales with the ordered moment as determined by neutron diffraction. Our new results indicate that below $x$= 0.01 the LMAF state rapidly weakens. The quasi-logarithmic temperature dependence of $\lambda_{LMAF}$ below $T_N$ shows that the internal magnetic dipolar fields measured at the $\mu^+$ localization site [28] grow only slowly with decreasing temperature.

In Fig.3 we show the magnetic and SC phase diagram for $x$< 0.012, highlighting our new μSR results. For $x$= 0.01, a $T_N$ value of 1.6±0.1 K has been extracted from zero-field μSR data taken on a polycrystal [15], while $T_N$= 1.7±0.1 K was obtained by single-crystal neutron-diffraction [21]. The SC ($T_c^+$) phase transition temperatures have been taken from Ref.26. Our new data for $T_N$ nicely follow the Doniach-diagram type behavior. From the data in Figs. 1 and 3, we can safely conclude that the LMAF phase line smoothly extrapolates to $T_N$= 0 at $x_{c,af} \approx 0.006$. Locating the magnetic QCP near $x$= 0.006 is consistent with the absence of any signal of the LMAF phase for $x$= 0.005, as was concluded from zero-field μSR measurements on a polycrystal down to 0.04 K [15], as well as from single-crystal neutron-diffraction data down to 0.1 K [21].

Our results show that *it is the LMAF phase which presents the magnetic instability in U(Pt,Pd)$_3$ and not SMAF*. This is consistent with recent transport measurements on polycrystalline U(Pt,Pd)$_3$ [29], which show clear deviations from Fermi-liquid behavior in the vicinity of $x_{c,af}$, as predicted for a QCP [30].

Inspecting the phase lines $T_N(x)$ and $T_c^+(x)$ plotted in Fig.3, we arrive at a most important conclusion, namely the critical concentration for the suppression of SC coincides with the critical concentration for the emergence of LMAF, $x_{c,s}=x_{c,af}=$ 0.006. The notion that the AF QCP is not found in pure UPt$_3$, but is reached upon doping, may end the longstanding debate of how an odd-parity SC state can arise [8,9], while the dominant fluctuations seem to be of AF nature [12]. In order to resolve this controversy, we propose that Pd doping leads to a shift of the spectral weight from FM to AF fluctuations. This idea is supported by the anomalously high rate of suppression of $T_c^+$ upon Pd substitution [26]. A shift of the spectral weight from FM to AF fluctuations is not uncommon near a QCP, where the many energy scales become comparable and competition between various phases becomes important. Indeed, inelastic neutron scattering experiments carried out on pure UPt$_3$ [31] show that the magnetic fluctuation spectrum is complex and has both AF and FM components.

The phase diagram shown in Fig.1, differs from the generic phase diagram, proposed for magnetically ordered pure heavy-fermion materials under pressure [3]. In these materials, the approach to the magnetic QCP at $T$=0 is circumvented by the occurrence of a SC ground state. $T_c$ is maximum near the critical pressure $p_{c,af}$, which has been interpreted as evidence for SC mediated by AF fluctuations. In the case of U(Pt,Pd)$_3$, however, $T_c \rightarrow 0$ at the QCP. This is naturally explained if *SC in UPt$_3$ is mediated by FM spin fluctuations,* which cannot coexist at any non-zero temperature with an ordered AF state. This is consistent with the notion that the SMAF state is fluctuating in time.

In conclusion, we have shown that the magnetic instability in U(Pt,Pd)$_3$ is due to the LMAF phase rather than SMAF. The U(Pt$_{1-x}$,Pd$_x$)$_3$ phase diagram has a critical point at $x \approx 0.006$ where unconventional SC is suppressed and LMAF emerges. The existence of this critical point provides new evidence for SC mediated by FM spin fluctuations. A complete understanding of the phase diagram and its quantum critical point might prove to be essential in further specifying the SC pairing mechanism. Measurements of the critical exponents of the thermal, magnetic and transport properties in the concomitant non-Fermi liquid regime are needed to identify the character of the magnetic fluctuations [30]. Moreover, the results may be relevant [32] to other strongly correlated systems, such as the high temperature superconductors, which have phase diagrams that exhibit a similar competition between SC and static AF order.

We thank K.S. Bedell and K.B. Blagoev for helpful discussions. This work was part of the research program of the Dutch "Stichting FOM". We acknowledge support through the EC/TMR program, the ESF/FERLIN program and the Petroleum Research Fund of the American Chemical Society.

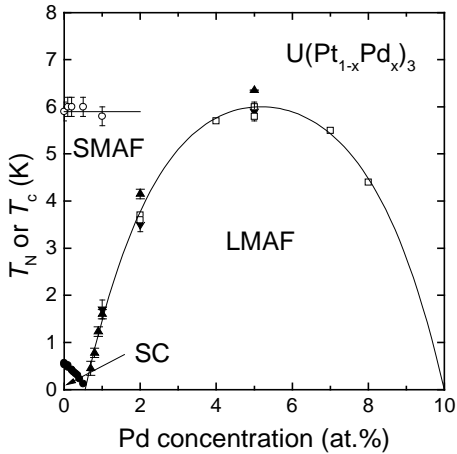

FIG. 1 Magnetic and superconducting phase diagram for U(Pt$_{1-x}$Pd$_x$)$_3$ alloys. SMAF = small-moment antiferromagnetic phase, LMAF = large-moment antiferromagnetic phase, SC = superconducting phase. Néel temperatures $T_N$ are measured by neutron-diffraction (o and ▼) [21], specific heat (□) [17,21] and μSR (▲) (Ref.15 and this work). Resistively determined superconducting transition temperatures $T_c^+$ (●) are taken from Ref. 26. The solid lines are to guide the eye.

FIG. 2 Temperature variation of the exponential relaxation rate, extracted from transverse field (100 G) μSR spectra using eq.(1), for U(Pt$_{1-x}$Pd$_x$)$_3$ with $x=$ 0.007, 0.008 and 0.009. The solid lines show the quasi-logarithmic increase of $\lambda_{LMAF}$ below $T_N$ and the temperature independent background $\lambda_{BG}$ above $T_N$.

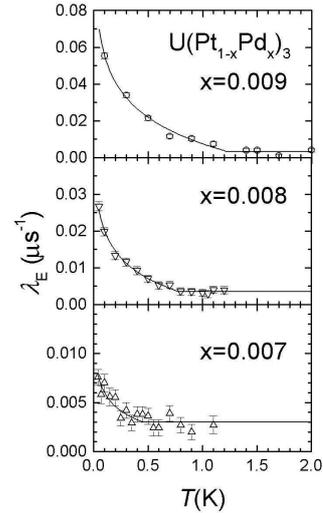

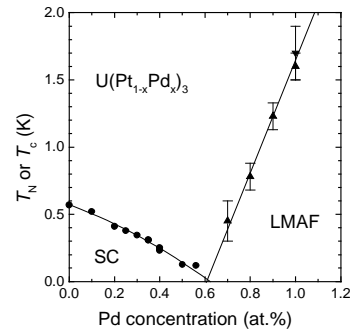

FIG. 3 Magnetic and superconducting phase diagram for U(Pt$_{1-x}$Pd$_x$)$_3$ alloys with $x< 0.012$. The meaning of the symbols is the same as in Fig.1. The solid lines serve to guide the eye.